\documentclass[nologo,11pt,a4paper,hidelinks]{ETHpaper}
\usepackage{graphicx, amsmath, amssymb,color,wasysym}
\usepackage{bm}
\usepackage[square,numbers,sort&compress]{natbib}
\usepackage{caption}

\begin{document}

\newcommand{\mean}[1]{\left\langle #1 \right\rangle} 
\newcommand{\abs}[1]{\left| #1 \right|} 
\newcommand{\ul}[1]{\underline{#1}}
\renewcommand{\epsilon}{\varepsilon} 
\newcommand{\eps}{\varepsilon} 
\renewcommand*{\=}{{\kern0.1em=\kern0.1em}}
\renewcommand*{\-}{{\kern0.1em-\kern0.1em}} 

\newcommand*{\+}{{\kern0.1em+\kern0.1em}}

\newcommand{\RA}{\Rightarrow}
\newcommand{\bbox}[1]{\mbox{\boldmath $#1$}}

\title{Should the government reward cooperation? \\ Insights from an
  agent-based model of wealth redistribution}

\titlealternative{Should the government reward cooperation? Insights from an
  agent-based model of wealth redistribution}

\author{Frank Schweitzer$^{0}$, Luca Verginer, Giacomo Vaccario}
\authoralternative{Frank Schweitzer, Luca Verginer, Giacomo Vaccario}

\address{Chair of Systems Design, ETH Zurich, Weinbergstrasse 58, 8092 Zurich, Switzerland}

\reference{(Submitted for publication)} 

\www{\url{http://www.sg.ethz.ch}}

\makeframing
\maketitle

\footnotetext{Corresponding author: \url{fschweitzer@ethz.ch}}

\begin{abstract}
  In our multi-agent model agents generate wealth from repeated interactions for which a prisoner's dilemma payoff matrix is assumed.
  Their gains are taxed by a government at a rate $\alpha$.
  The resulting budget is spent to cover administrative costs and to 
  pay a bonus to cooperative agents, which can be identified correctly only with a probability $p$.
  Agents decide at each time step to choose either cooperation or defection based on different information.
  In the local scenario, they compare their potential gains from both strategies.
  In the global scenario, they compare the gains of the cooperative and defective subpopulations.
  We derive analytical expressions for the critical bonus needed to make cooperation as attractive as defection.
  We show that for the local scenario the government can establish only a medium level of cooperation, because the critical bonus increases with the level of cooperation. 
  In the global scenario instead full cooperation can be achieved once the cold-start problem is solved, because the critical bonus decreases with the level of cooperation.
  This allows to lower the tax rate, while maintaining high cooperation.
  
\end{abstract}

\date{\today}

\section{Introduction}

A high level of cooperation is generally desired from the perspective of a social planner.
At the same time, such a systemic state is hard to achieve, and even more difficult to \emph{maintain}.
The reasons for this are discussed in game theory from a theoretical point of view, and in behavioral economics in an experimental and psychological context.
The bottom line is that non-cooperative behavior, i.e. defection, often leads to higher gains, whereas cooperative behavior implies a cost.
Therefore, selfish and rational agents would choose defection.

Over the last seven decades, researchers have developed various proposals of how to solve this dilemma, which we cannot review or even list here.
The punishment of defective behavior is one of these solutions.
We know that punishment is costly, but useful to keep cooperation high.
It is also known that partial punishment  applied only during some periods or on a fraction of the population is already sufficient to maintain cooperation~\citep{boyd1992punishment,fehr2002altruistic, fowler2005altruistic}.

However, when agents have the choice of costly punishments, the rational strategy is to avoid the costs incurred and to leave the punishment to other agents.
This is not confirmed by experiments, but is still a common explanation for the decreasing level of cooperation~\citep{fehr2003nature}. 
In the end, costly punishments are not a sustainable solution to keep cooperation high, because of \emph{second-order effects}.
Similar to choosing defection if others cooperate, it pays off to choose not to punish if others punish.
One way out of this dilemma is to assign the task of punishment to a central authority, such as a government.

Nowadays, it becomes increasingly popular even on a governmental level to replace punishments which have a negative connotation with bonuses for those who follow certain rules.
This is an indirect punishment, because those who do not follow these rules, will not receive a bonus.
From a psychological perspective positive incentives work much better than negative deterrents, although it was also found that they do not work perfectly ~\citep{hauert2010replicator,moisan2018not,pulford2018incentive,kajackaite2017incentives}.

In this paper, we develop an agent-based model that allows to systematically investigate the role of bonuses paid for cooperative behavior.
We choose a game theoretical setting because it already quantifies the gains from cooperative and defective behavior.
Our main idea is to let the government collect taxes on these gains, which allows to redistribute the wealth accumulated this way.
Wealth redistribution models have been already discussed with respect to agents' interactions \citep{slanina2004inelastically,during2008kinetic} or assuming a central authority \citep{Lorenz2013,atkinson1971capital}.

Compared to these rather mathematical models our approach is simpler, but still analytically tractable.
It has more in common with public good games (PGG), where agents can contribute to a common pool that returns benefits to participants~\citep{mancur1965logic,yamagishi1986provision,Ledyard_2020,tessone2013c}.
Different from these type of games, in our model agents are obliged to contribute to the common good with an amount that depends on their gains.
This is ensured by the taxation from the central authority, i.e. the government.

The redistribution mechanism in our model shall influence agents in their decision to either cooperate or defect.
Precisely, from the collected taxes the government pays a bonus only to cooperators, once they have been identified as such.
With respect to these external incentives, our model bears some similarities to \emph{nudging}~\citep{thaler2009nudge} in that the central authority does not forbid a specific behavior.
But, different from a nudge, we not only use the bonuses as economic incentives to change the  behavior of the agent, but also different types of information.
In the local information regime agents decide about their cooperative or defective 
behavior by comparing their potential gains, whereas in the global information regime they compare the current gains of the cooperative and defective subpopulations.

The global information regime introduces a non-linear feedback between the fraction of cooperators in the system and the total amount of bonuses needed to reward further cooperation.
Because of this the critical bonus decreases with the fraction of cooperators, but the level of cooperation continues to increase.
This eventually leads to a scenario of \emph{sustainable cooperation}, where the government can choose low tax rates because bonuses are no longer needed. 

These interesting findings are obtained analytically and further confirmed by means of agent-based simulations.
In the following, we separately introduce the agent and the governmental perspectives.
We then derive expressions for the critical bonus needed to obtain cooperation.
The drawbacks resulting from the local information regime will motivate the proposal for the global information regime, which leads to the promising results.

\section{The agent perspective}
\label{sec:agent-model}

\subsection{Wealth, strategies, bonuses and taxes}
\label{sec:agent-wealth}

In the following, we consider a number of agents $i=1,...,N$. 
Each agent is characterized by an internal state variable, $s_{i}(t)$, which indicates a cooperative ($s=1$) or a defective ($s=0$) \emph{strategy}, as generally considered in game theory.
Further, each agent possesses some \emph{wealth}, $w_{i}(t)$, taken in a very general sense, which changes over time and can also accumulate.

A gain of wealth, $g_{i}(s_{i},t)$, in a given time step can come from two sources, (i) interactions with other agents $j$ which are described by an interaction term $\mathcal{F}_{ij}(t)$ discussed in a separate section, and (ii) a \emph{bonus} $b_{i}(s_{i},t)$ paid by a central authority, called the government in the following.
\begin{align}
  \label{eq:40}
  g_{i}(s_{i},t)= \sum\nolimits_{j} \mathcal{F}_{ij}(t) + b_{i}(s_{i},t)
\end{align}
In our paper the bonus payment depends on how the agent ``behaves'' in a very general sense.
Agents can, for instance, be rewarded by the government if they comply to social norms, respect laws, protect the environment or provide community services. 
Utilizing a game theoretical setting, we call such behavior \emph{cooperative}.
The opposite would be a \emph{defective} behavior, e.g. to not comply to a social norm or to contribute to a common good. 
We assume that bonuses are \emph{only} paid to agents with a cooperative behavior, i.e. if $s_{i}(t)=1$.
We can express the bonus as follows:
\begin{align}
  \label{eq:2}
    b_{i}(s_{i},t)= b(t) \,\Theta[z_{i}(t)]
\end{align}
$\Theta[x]$ is the Heaviside function with $\Theta[z]=1$ if $z\geq 0$ and $\Theta[z]=0$ if $z<0$.
Hence, to make sure that for cooperators $b_{i}$ is positive, the variable $z_{i}(t)$ has to be equal or larger than zero.
This is guaranteed e.g. by choosing $z_{i}(t)=2s_{i}(t)-1$, which returns positive values for cooperators and negative ones for defectors. 
$b(t)$ is the amount paid to each cooperator.
Its time dependence is discussed in the next section.

One could argue that, in order to pay the bonus only to cooperators, $b_{i}(s_{i},t)=b(t)s_{i}(t)$ would be sufficient.
It is, indeed, as long as there are no mistakes in identifying cooperators and defectors. 
That means the government has not only complete and perfect knowledge about the strategies of all agents, it also makes no administrative mistakes.
A more realistic setting should consider that there is only a (large) probability $p\in(0,1)$ that defectors (and cooperators) are identified correctly,  whereas with a (small) probability $(1-p)$ a cooperator is wrongly identified as a defector and a defector as a cooperator.
This is reflected in $z_{i}(t)$ which becomes a stochastic variable, to assign bonuses to agents from the governmental perspective.
The dynamics for $z_{i}(t)$ reads:
\begin{align}
  \label{eq:3}
z_{i}(t)=\left[2s_{i}(t)-1\right] \mathcal{B}(k,t) \;;\quad \mathcal{B}(k,t) \in \{-1,+1\} \;;\quad \mathcal{B}(k,t) \sim p^{\frac{1+k}{2}} (1-p)^{\frac{1-k}{2}}
\end{align}
where $\mathcal{B}(k,t)$ is a random Boolean variable which follows a Bernoulli distribution.
That means, it takes the value $k=+1$ with a probability $p$ and the value $-1$ with the probability $(1-p)$. 

Eventually, we consider that at each time step the individual gain $g_{i}(s_{i},t)$ is \emph{taxed} by the government at a tax rate $\alpha$, which is constant and the same for all agents.
This tax can be interpreted as an \emph{income tax}, because it considers gains, or incomes, only from the current time interval $\Delta t$.
With this, we propose the following discrete dynamics for the wealth of agent $i$, with $\Delta t=1$:
\begin{equation}
  \label{eq:wi}
  w_{i}(s_{i},t+1)= w_{i}(s_{i},t) + (1-\alpha)  g_{i}(s_{i},t)
 \end{equation}
This is known as \emph{proportional tax} scenario, because it is proportional to the income of each agent.  
Other scenarios of tax collections have been discussed in \citep{Lorenz2013}.

Alternatively, one could assume that instead of the gain the \emph{wealth} of an agent is subject to taxes. This would modify Eqn.~\eqref{eq:wi} into:
\begin{equation}
  \label{eq:wi2}
  w_{i}(s_{i},t+1)= (1-\alpha) w_{i}(s_{i},t) +  g_{i}(s_{i},t)
\end{equation}
There are two arguments against this choice.
First of all, in our model individual wealth can be \emph{only} accumulated in one way, namely through the gain.
So, from an \emph{economic} perspective, it is sufficient to tax this wealth contribution instantaneously, like an annual income tax.
Taxing the (accumulated) wealth instead implies that gains from previous time steps will be implicitly taxed again, at a discounted rate.
Because of this Eqn.~\eqref{eq:wi2} becomes in fact a \emph{recursive} equation.
This problem could be mitigated by assuming stationarity, $ w_{i}(s_{i},t+1)= w_{i}(s_{i},t)$, which however implies that at every time step each gain equals the taxes paid by an agent. 
Therefore, in the following, we will consider Eqn.~\eqref{eq:wi}. 

We note that the gain is always greater or equal to zero, $g_{i}(s_{i},t)\geq 0$.
This implies that individual wealth can only grow according to Eqn.~\eqref{eq:wi}.
It could also stay constant, in the worst case, $g_{i}(s_{i},t)=0$. To specify the two contributions to this positive gain $g_{i}(s_{i},t)$, we need to discuss strategic interactions of agents and the value of the bonuses paid to cooperative agents.

\subsection{Strategic interactions of agents}
\label{sec:strat-inter-agents}

Eq.~\eqref{eq:wi} considers the opportunity to increase wealth via dyadic interactions between two agents $i$ and $j$, expressed in the term $\mathcal{F}_{ij}(t)$.
Using the game theoretical setting, we assume that this interaction is described by the so-called \emph{prisoner's dilemma}.  
This means, if two agents $i,j$ interact, their resulting payoff depends on both $s_{i}$, $s_{j}$ and is, in general,  \emph{asymmetric}: $A_{i}[s_{i},s_{j}]\neq A_{j}[s_{j},s_{i}]$.
Hence, the classical payoff matrix reads as:
\begin{align}
  \label{eq:4}
  \begin{tabular}{c|c|c}
	& $s_{j}=1$ & $s_{j}=0$ \\ \hline
	$s_{i}=1$ & $A_{i} [1,1] \ ; \  A_{j}[1,1]$ & $ A_{i}[1,0] \ ; \  A_{j}[1,0]$ \\ \hline
	$s_{i}=0$ & $A_{i}[0,1] \ ; \  A_{j}[0,1]$ & $ A_{i}[0,0] \ ; \  A_{j}[0,0]$ \\ \hline
	\end{tabular}
 \quad =\quad
	\begin{tabular}{c|c|c}
	& C  & D \\ \hline
	C & R/R & S/T \\ \hline
	D & T/S & P/P \\ \hline
	\end{tabular}
\end{align}
The dilemma arises from the fact that
\begin{align}
  \label{eq:5}
  T>R>P>S, \;; \quad 2\,R > S+T
\end{align}
i.e. the rational choice would be always to defect because, no matter what the counterparty chooses, the payoff for defectors would be always higher.
However, if both agents cooperate their joint payoff is larger, i.e. the social welfare, as the sum over all payoffs, is maximized.

In our model, we assume that during each time step one agent, with a \emph{fixed strategy} $s_{i}(t)$, plays $n_{g}$ of the described so-called 2-person games with \emph{different} agents.
This means, the counterparty is chosen \emph{randomly} from the population of $N-1$ agents such that no agent is not chosen twice during the $n_{g}$ interactions with a specific agent. This ensures that during each time step each agent receives $n_{g}$ payoffs from different agents. 
We take these payoffs as a proxy for $\mathcal{F}_{ij}$, i.e. 
\begin{align}
  \sum_{j\subseteq N, j=1}^{n_{g}} \mathcal{F}_{ij}(t)= \sum_{j\subseteq N, j=1}^{n_{g}} A_{i}[s_{i}(t),s_{j}(t)]
  \label{eq:8}
\end{align}
During the $n_{g}$ interactions at time $t$, agent $i$ may meet $n^{1}_{i}(t)$ times a cooperator and $n^{0}_{i}(t)=n_{g}-n^{1}_{i}(t)$ times a defector.
This allows to express the cumulative payoff as 
\begin{align}
  \hat{A}_{i}[s_{i},n_{i}^{1}]= \delta_{1,s_{i}}\left[n_{i}^{1}R+(n_{g}-n_{i}^{1})S\right] 
  + \delta_{0,s_{i}}\left[n_{i}^{1}T+(n_{g}-n_{i}^{1})P\right]
  \label{eq:9}
\end{align}
where the Kronecker delta $\delta_{s,s_{i}}=1$ if $s=s_{i}$.

We can further define, with respect to agent $i$, an \emph{individual frequency of cooperators}  $f_{i}^{1}(t)\equiv f_{i}(t)=n^{1}_{i}(t)/n_{g}$.  
In our setup, however, interacting agents are sampled uniformly at randomly from the population.
Therefore, the individual frequencies can be approximated by  the global frequency of cooperators, $f_{i}(t)\equiv f(t)=N_{c}(t)/N$, where $N_{c}(t)$ denotes the total number of cooperators and $N_{d}(t)=N-N_{c}(t)=N[1-f(t)]$ the total number of defectors. 
This so-called mean-field limit is applied in the following. 

Using $f(t)$, we introduce the \emph{relative payoff}
\begin{align}
\hat{A}_{i}[s_{i},n_{i}^{1}]= n_{g}\,  a_{i}[s_{i},f]
  \label{eq:10}
\end{align}
which, with reference to the payoff matrix, Eq.~\eqref{eq:4}, reads as
\begin{align}
  \label{eq:11}
  {a}_{i}\left[1,f\right] =& {a}_{i}\left[s_{i}=1,f\right] = R\ + S\ \left [1-f\right]
  \nonumber \\
  {a}_{i}\left[0,f\right] =& {a}_{i}\left[s_{i}=0,f(t)\right] =  T\ f + P \ \left [1-f\right]
  \end{align}

\subsection{Strategic decisions: Local information}
\label{sec:strategic-decisions}

We follow a stochastic approach, i.e. agent $i$ uses a strategy $s_{i}(t)$ only with a certain probability $q_{i}(s_{i},t)$.
Taking $s_{i}=1$, the dynamics is then described by the master equation:
\begin{align}
  \label{eq:18}
    q_{i}(1,t+1)-\,q_{i}(1,t)
  =-q(0|1,\rho_{i})\,q_{i}(1,t) +q(1|0,\rho_{i})\,\left[1-q_{i}(1,t) \right] \approxeq \frac{dq_{i}(1,t)}{dt} 
\end{align}
$q(0|1,\rho_{i})$ and $q(1|0,\rho_{i})$ are the transition probabilities to switch between the two strategies $s_{i}=0$ and $s_{i}=1$,
i.e. they model the \emph{decision} of agent $i$  based on the current information encoded in $\rho_{i}$.

In this paper, we consider two different specifications for $\rho_{i}$, \emph{local} and \emph{global} information.  
Local information refers to the fact that the agent is able to calculate in advance the gain to be expected from a given strategy $s_{i}$.
This is a classical assumption of game theory.
The agent then compares the \emph{expected gains} and decides to choose the strategy that will return the higher gain in the next time step, $g_{i}(s_{i},t+1)$, with a higher \emph{probability}.
That means the term \emph{local} indicates that the agent only considers its individual perspective. 

In our model, we assume that agents, when deciding about their strategy, do not only take the payoff from interactions with other agents into account, but also the bonus they may receive from the government and the taxes they have to pay on that.
Hence, with
\begin{align}
  g_{i}(s_{i},t+1) = n_{g}\, a_{i}[s_{i}(t),f(t)] + b_{i}(s_{i},t)
  \label{eq:12}
\end{align}
we define the transition probabilities as follows:
\begin{align}
  \label{eq:13}
q(s_{i}|(1-s_{i}),f)
  = \frac{\exp{\{\beta \ g_{i}(s_{i},f)\}}}{
\exp{\{\beta \ g_{i}(s_{i},f)\}}+\exp{\{\beta \ g_{i}(1-s_{i},f)\}}}
\end{align}
The parameter $\beta$ denotes the level of randomness and determines how sensitive agents are with respect to wealth differences.
$\beta \to 0$ implies random choices, $\beta \to \infty$ means that already small wealth differences result in strategy changes.

This strategic behavior has two different uncertainties for an agent:
(i) there is a small probability $(1-p)$ that the agent is not rewarded if it cooperates, or still rewarded if it defects,
(ii) because the counterparty for an interaction is randomly chosen, this results in a mix of cooperators and defectors during the $n_{g}$ interactions.
Further, the probability to meet a cooperator or a defector is changing over time as agents change their strategy.

\subsection{Critical bonus level}
\label{sec:odds-ratio}

Already at this point we are able to discuss critical parameter constellations that may change the outcome of the dynamics.
Given the master eq.~\eqref{eq:18}, stationarity, i.e. $q_{i}(1,t+1)-\,q_{i}(1,t)=0$, implies the so-called \emph{detailed balance condition}
    \begin{align}\label{eq:15}
      \frac{q_{i}(1)}{1-q_{i}(1)}=\frac{q(1|0,f)}{q(0|1,f)}
    \end{align}
Using the transition probabilities, Eq.~\eqref{eq:13}, we can calculate from the stationary solution the Logit or log odds ratio: 
\begin{align}
   \mathcal{G}\left[q(1,f)\right] \equiv \ln \frac{q(1,f)}{1-q(1,f)} = 
      \beta \left[g_{i}(s_{i}=1,f) -g_{i}(s_{i}=0,f)\right]
\label{eq:14}
\end{align}
\texttt{logit}($s_{i}=1)>0$ tells under what conditions it is more likely that the agent chooses cooperation, $s_{i}=1$, instead of defection.
With Eq.~\eqref{eq:12}, this results in:
\begin{align}
  \label{eq:16}
   n_{g}\, a_{i}[s_{i}=1,f] + b_{i}(s_{i}=1) >  n_{g}\, a_{i}[s_{i}=0,f] + b_{i}(s_{i}=0)
\end{align}
The respective payoffs are given by Eq.~\eqref{eq:11} and the expected bonuses by Eq. \eqref{eq:2}.
To correctly determine $b(t)$, we have to consider the probability $p$ that bonuses are wrongly assigned.
Because such events are independent, $p$ already gives the fraction of correctly identified cooperators, whereas $(1-p)$ gives the fraction of defectors that will receive a bonus by mistake.
We define $b_{i}(s_{i}=1)=p b(t)$ and $b_{i}(s_{i}=0)=(1-p) b(t)$, where $b(t)$ is the equal bonus paid to all agents identified as cooperators, according to Eq.~\eqref{eq:2}.
With this, we find as the critical condition for an agent to choose cooperation:
\begin{align}
  \label{eq:17}
  R f+ S \left [1-f\right] + \frac{p}{n_{g}}\, b(t) > T f + P  \left [1-f\right] + \frac{(1-p)}{n_{g}} b(t) 
\end{align}
This leads to the \emph{critical value} of the bonus that has to be reached at minimum, in order to make cooperation attractive:
\begin{align}
  \label{eq:20}
  b(t) >  b^{\mathrm{crit}}(f,p)= \frac{n_{g}}{2p -1} \big\{(P-S) + f \left[(T-R)-(P-S)\right] \big\} 
\end{align}
Inserting the canonical payoff values $\{T,R,P,S\}=\{5,3,1,0\}$ we obtain:
\begin{align}
  \label{eq:21}
   b^{\mathrm{crit}}(f,p)= \frac{n_{g}[1+ f]}{2p -1}  \end{align}
The critical bonus $b^{\mathrm{crit}}(f,p)$ in Eq.~\eqref{eq:21} has a clear interpretation: it is the bonus sufficient to compensate the payoff difference of a cooperator and a defector in the same situation, i.e. cooperation becomes as attractive as defection.

\begin{figure}[htbp]
  \centering
  \includegraphics[width=0.45\textwidth]{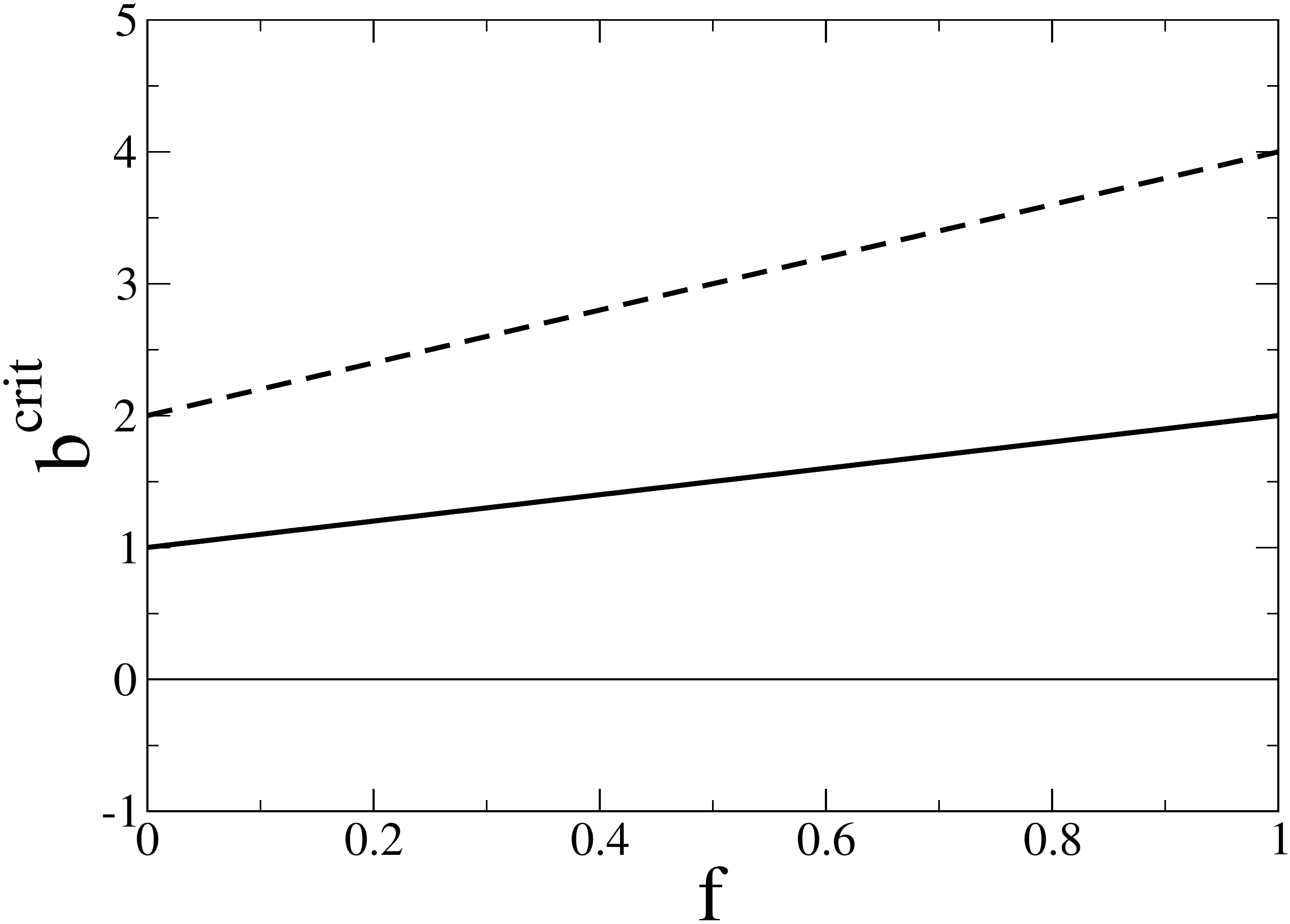}
  \caption{Critical bonus level, Eqn.~\eqref{eq:21} dependent on the fraction of cooperators, $f$. Solid line: $p$=1, dashed line: $p$=0.75.}
  \label{fig:bonus1}
\end{figure}
From Figure~\ref{fig:bonus1} 
we see that the higher the level of cooperation $f$, the higher the bonus has to be.
This reflects the underlying assumption of game theory, namely that it becomes more attractive to choose defection if an agent is surrounded by cooperators.
So, the bonus in fact reflects the higher incentive needed to \emph{not} exploit such a situation.
If the agent interacts only with defectors ($f=0$), then ${b}^{\mathrm{crit}}\propto (P-S)=1$ is sufficient to make cooperation an equally attractive option.
In contrast, if it interacts only with cooperators ($f=1$), then ${b}^{\mathrm{crit}}\propto (T-R)=2$ is needed to not exploit cooperation.
This bonus has to be paid $n_{g}$ times, i.e. for every interaction, and hence we have $n_g$ as factor in Eq.~\eqref{eq:21}.

The critical bonus level further depends on $p$, the accuracy in detecting cooperators and defectors correctly.
For $p\to 0.5$, both defectors and cooperators have the same probability to receive a bonus.
In this case, $b^{\mathrm{crit}}\to\infty$ and hence, the payoff differences can no longer be compensated, and the nudging mechanism fails.
If $p=1$, we reach, for a given $f$, the lowest value for the critical bonus. 
If $p$ drops by 25\%, the critical bonus already doubles.

\section{The government perspective} 
\label{sec:government}

\subsection{Wealth and  taxes}
\label{sec:wealth-taxes}

The second important player in our model is the government.
Similar to the agents, the government also possesses some wealth, $W(t)$. 
Governmental wealth can be only accumulated via tax collection.
These taxes are used to pay the bonuses to those agents identified as cooperators.
Hence, for the accumulated wealth of the government, we propose the following dynamics:
\begin{align}
  \label{eq:22}
  W(t+1)=W(t) + \alpha G(t) - b(t) N_{b}(t) \; ;
  \quad G(t)= \sum\nolimits_{i} g_{i}(s_{i},t)
\end{align}
$N_{b}$ is the number of bonuses paid by the government.  
We recall that the government can identify cooperators only with a given probability $p$.
With a probability $(1-p)$ also defectors will receive a bonus, by mistake.
Therefore, from the governmental perspective, instead of the correct number of cooperators $N_{c}(t)$, the government has identified a number
\begin{align}
  \label{eq:28}
  N_b(t) = p \ N_c(t) + (1-p) N_{d}(t) = N\left\{p f(t) + (1-p) [1-f(t)]\right\}
\end{align}
of agents that actually have received the bonus ${b}(t)$.

If the government wants  to accumulate wealth, at least for some time periods, it has to ensure that
\begin{align}
  \label{eq:35}
 W(t+1)-W(t)= \Delta W(t) = \alpha G(t) - N b(t) \left\{p f + (1-p) [1-f]\right\} \geq 0
\end{align}
In an ideal economy without credit $\Delta W(t)$, the wealth generated in one time step, defines the annual governmental budget, i.e. the upper limit of what can be spent by the government.
This budget, first and foremost, has cover administrative tasks.
Specifically, at each time step the government has (i) to collect taxes from all agents, (ii) to identify cooperating agents with a probability $p$, (iii) to pay bonuses to those identified as cooperators.
These activities are costly, thus we consider an administrative cost $C$, and split the  budget as follows: $\Delta W(t)=C(t)+S(t)$. 
$S(t)$ denotes the governmental \emph{savings}, i.e. the part of the budget not used to pay the administrative costs. 
In a subsequent paper, we discuss different assumptions of how the administrative costs change e.g. with the level of cooperation, $f$, and how they depend on the accuracy $p$ to identify cooperators.
Here, we simply assume that $C$ and $S$ are two constants equal to zero.
This simpler case allows to study a relation between the budget $\Delta W$, the tax level $\alpha$, the level of cooperation $f$, and the accuracy $p$  to identify cooperators.

The total taxes collected by the government at each time step result directly from Eqn.~\eqref{eq:12}: 
\begin{align}
  \label{eq:31}
  \alpha G(t) = \alpha \sum_{i=1}^{N} g_{i}(s_{i},t)= & \alpha  {N f} \left\{n_{g} \big[  R f+ S \left [1-f\right]\big] + {p}\, {b}\right\} \nonumber \\
  + & \alpha {N [1-f]} \left\{ n_{g} \big[  T f+ P \left [1-f\right]\big] + {(1-p)}\, {b} \right\}
\end{align}
With this expression, we can determine  parameter constellations under which the condition of Eqn.~\eqref{eq:35} can be met.
This is discussed in the following section.

\subsection{Governmental budget}
\label{sec:administrative-costs}

Let us first calculate Eqn.~\eqref{eq:31}, using the canonical payoff values $\{T,R,P,S\}=\{5,3,1,0\}$:
\begin{align}
  \label{eq:36}
  \alpha G(t)= \alpha N n_{g}\left\{-f^{2}+3f+1\right\}+ \alpha N b(t)
  \left\{pf+(1-p)[1-f]\right\} 
\end{align}
The second term reflects the fact that bonuses paid by the government are also subject to taxes, and this way are partly re-collected. 
Hence, effectively the government only distributes a share $(1-\alpha)$ to the agents and we find for the upper limit of the available budget:
\begin{align}
  \label{eq:37}
  \Delta W(t) \leq \alpha N n_{g}\left\{-f^{2}+3f+1\right\} - 
  (1-\alpha) N b(t)
  \left\{pf+(1-p)[1-f]\right\} 
\end{align}
We already know from Eqn.~\eqref{eq:20} that $b(t)$ has to reach a minimum critical level $b^{\mathrm{crit}}$ in order to be effective, i.e. to let 
differences between the wealth of cooperators and defectors vanish.
Thus, we assume $b(t)=b^{\mathrm{crit}}$, Eqn.~\eqref{eq:21}, which results finally in:
\begin{align}
  \label{eq:38}
 \Delta W \leq \alpha N n_{g}\left\{-f^{2}+3f+1\right\}
  - (1-\alpha) N n_{g} \frac{1+f}{2p-1}\left\{p f + (1-p) [1-f]\right\}  
\end{align}
which simplifies to:
\begin{align}
  \label{eq:39}
  {\delta}(f,\alpha,p)=\frac{\Delta W}{N n_{g}} \leq \left\{- f^{2} + 3\alpha f +\alpha\right\} - \frac{1-\alpha}{2p-1}\left\{pf+(1-p)\right\}
\end{align}

\begin{figure}[htbp]
  \centering
   \includegraphics[width=0.45\textwidth]{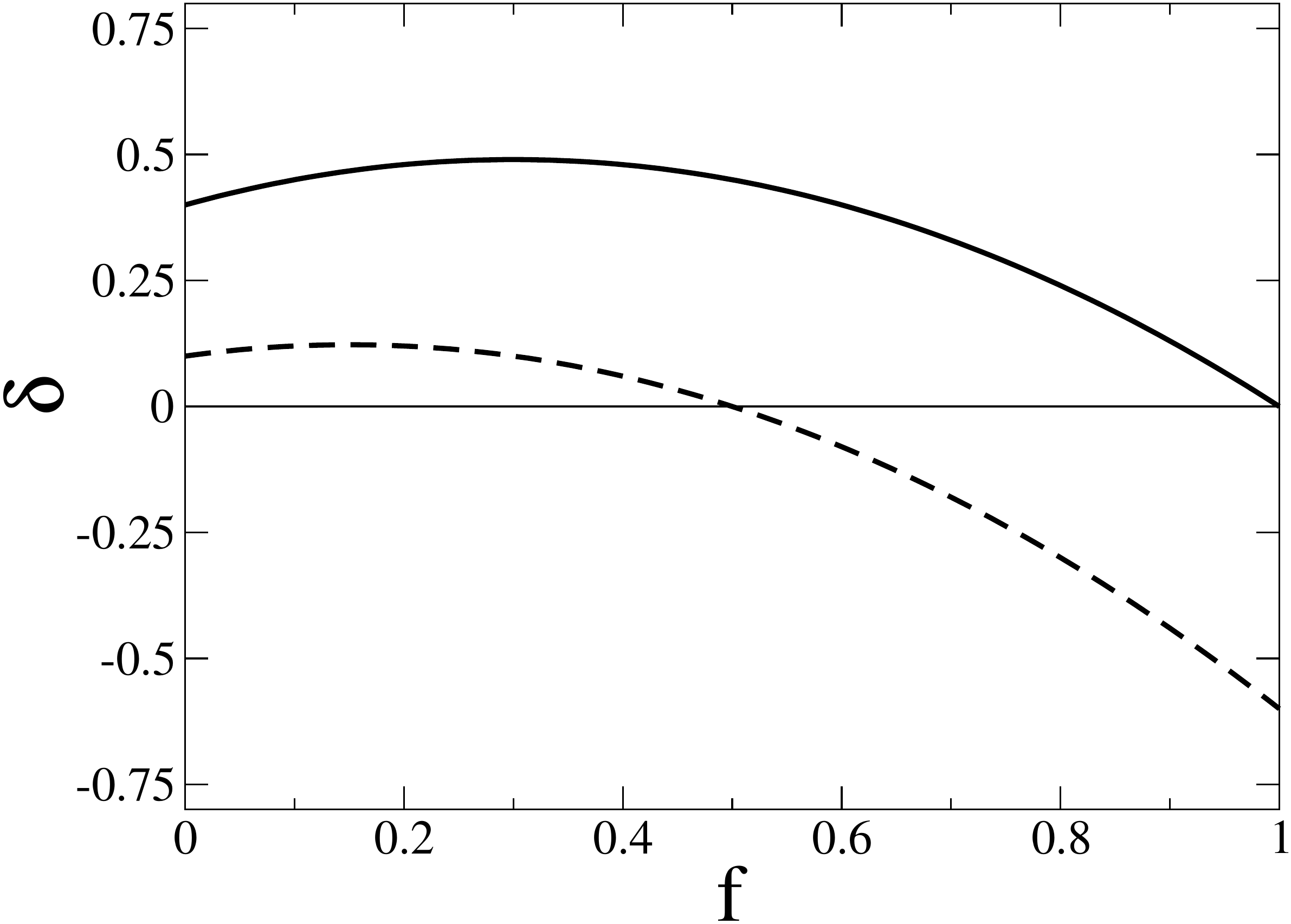}(a)
  \hfill
  \includegraphics[width=0.45\textwidth]{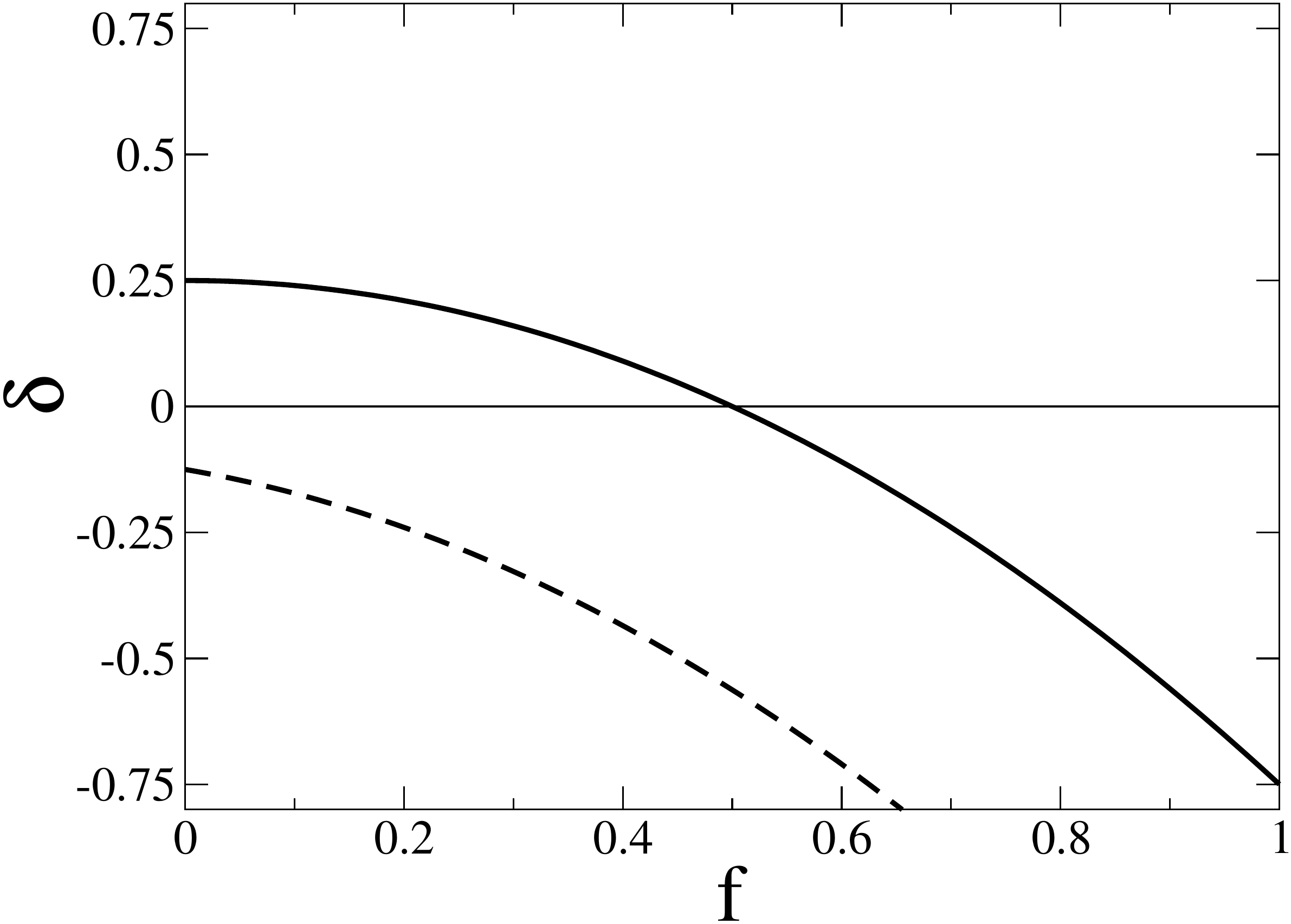}(b)
  \caption{Budget level, Eqn.~\eqref{eq:39} dependent on the fraction of cooperators, $f$. Solid line: $p$=1, dashed line: $p$=0.75. Different tax levels: (a) $\alpha$=0.4, (b) $\alpha$=0.25.} 
  \label{fig:costs1}
\end{figure}

If we assume $p=1$, this results in:
\begin{align}
  \label{eq:41}
  {\delta}(f,\alpha,p=1)\leq -f^{2} + (4\alpha-1) f +\alpha
\end{align}
This already allows to conclude that the government can cover its administrative costs \emph{and} pay the critical bonus for \emph{all levels} of cooperation, including $f=1$, only if $5\alpha-2\geq 0$, or $\alpha\geq 0.4$.
As a comparison in Figure~\ref{fig:costs1} shows, 
for lower tax rates bonuses can only be afforded for lower levels of $f$.
In other words, the government can push the level of cooperation only to a critical fraction $f^{\mathrm{crit}}$ that results from ${\delta}=0$ (assuming that it drops all own costs)
\begin{align}
  \label{eq:42}
  f^{\mathrm{crit}}(\alpha,p=1)=\frac{(4\alpha-1)}{2} + \frac{1}{2} \sqrt{(4\alpha-1)^{2}+4\alpha}
\end{align}
For instance, $\alpha=0.25$ results in $f^{\mathrm{crit}}=0.5$,  $\alpha=0.3$ gives $f^{\mathrm{crit}}=0.64$.

If we consider $p<1$, it is already obvious from Eqn.~\eqref{eq:39} that the terms involving $p$ are all positive, therefore ${\delta}$ would be only lowered.
For example, $\alpha=0.4$ and $p=0.75$, would result in a critical value $f^{\mathrm{crit}}(\alpha,p)=0.5$, and for $\alpha=0.25$ and $p=0.75$ no bonuses can be paid, as Figure~\ref{fig:costs1} shows. 
In other words, to compensate for the ``mistakes'' made by the government in identifying cooperators, the tax rate needs to be further increased, to reach the same level of cooperation.

\begin{figure}[htbp]
  \centering
  \includegraphics[width=0.45\textwidth]{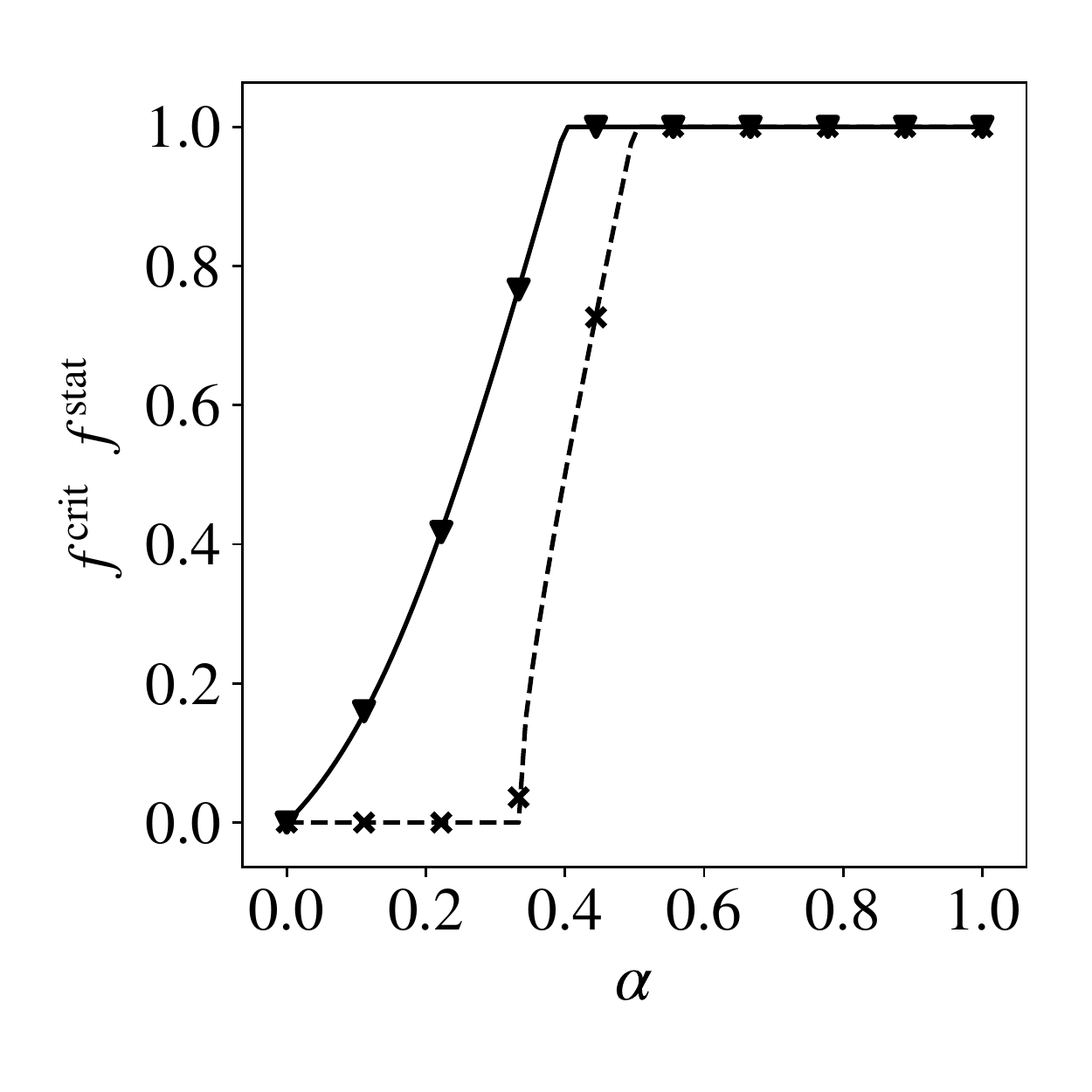}
  \caption{Comparison of $f^{\mathrm{crit}}$ (lines), Eqn.\ref{eq:42}, and the stationary level of cooperation $f^{\mathrm{stat}}$ (symbols) obtained from agent-based simulations for different levels of the tax rate $\alpha$.
  $p$=1 (continuous line, triangles), $p$=0.75 (dashed line, crosses).  
  }
  \label{fig:sim1}
\end{figure}

To validate the analytical solution of Eqn.~\eqref{eq:42}, we performed stochastic simulations of our agent-based model.
Specifically, we simulate the master Eqn.~\eqref{eq:18} with the transition rates, Eqn.~\eqref{eq:13} and the gains, Eqn.~\eqref{eq:12}, assuming $n_{g}(N-1)/2$ different interactions, chosen randomly without replacement.
We then run the simulations with a given set of parameters until the fraction of cooperators $f(t)$ converges to a stationary value, $f^{\mathrm{stat}}$.
This value is then compared with the analytical prediction of the expected $f^{\mathrm{crit}}$, Eqn.~\eqref{eq:42}.
In Figure~\ref{fig:sim1}, this comparison is presented as a function of the tax rate $\alpha$ for two different values of $p$.
We find a perfect match between the analytical and the simulation results.

A last discussion answers the question, for which level of cooperation the government would have the highest possible budget ${\delta}(f,\alpha,p)$ available.
We start from the nonlinear Eqn.~\eqref{eq:39} and take the first derivative, to obtain the optimal condition
\begin{align}
  \label{eq:23}
  f^{\mathrm{opt}}(\alpha,p)= \frac{1}{2}\left[3\alpha  - \frac{p(1-\alpha)}{2p-1}\right]
\end{align}
Thus, for $p=1$ and $\alpha=0.4$ we obtain $f^{\mathrm{opt}}(\alpha,p)=0.3$.
Hence, for $0\leq f\leq f^{\mathrm{opt}}$, it pays off for the government to pay bonuses, because this not only increases the level of cooperation, but also the budget for the government. 

For $f^{\mathrm{opt}}\leq f\leq f^{\mathrm{crit}}$, the government is still able to pay the critical bonus, this way further increasing the level of cooperation, albeit its own budget starts to decrease with increasing $f$, until it vanishes for $f=f^{\mathrm{crit}}$. 
Therefore, from the perspective of the governmental budget, $f=f^{\mathrm{opt}}$ would be the preferred level of cooperation.

\section{Strategic interaction with global information}
\label{sec:strat-inter-with}

\subsection{Critical bonus level}
\label{sec:critical-bonus-level-1}

The previous results are interesting  because they explain (i) under what conditions it pays off for the government to reward cooperation, and (ii) which level of cooperation could be achieved by such a bonus system.
At the same time, the results are also ``boring'' because they reflect basic insights already known from classical game theory.
If agents choose their strategy by comparing their potential gains, as we have assumed in Eqn.~\eqref{eq:13}, in a classical prisoner's dilemma they will always choose defection because it gives the higher individual gain in every possible interaction.
The bonus paid by the government to reward cooperation is precisely the difference between these potential gains, to make cooperation as attractive as defection.

Thus, the government has ``bought'' cooperative behavior, which sustains as long as bonuses are paid.
The only interesting insight was that, dependent on the tax level, the government cannot ``buy'' any high level of cooperation, because it cannot afford to pay the bonuses.
These bonuses have to be much higher in cooperative environments because the temptation to defect in this situation is also much higher.
Apart from quantifying these relations, the conclusions are not really new.

This leaves us with the important question how to change the model such that \emph{non-trivial} results are obtained.
We refrain from applying known remedies to boost the level of cooperation, such as considering local or iterated interactions \citep{hauert02:_effec_games,Nowak_1994,schweitzer2005,fs-lb-acs-02}, or introducing additional mechanisms such as social herding \citep{Glance:94,pavlin2013} or \emph{migration} \citep{jiang2010role,schweitzer2012b}. 
Instead, we change the model only at one point, namely in the information the agent takes into account when deciding about its strategy.

So far, the agent only considers the ``local'' or individual perspective, by comparing its potential gains, reflected in Eqn.~\eqref{eq:13}.
Now we assume that the agent compares the global payoffs of the two subpopulations,  defectors and cooperators.
This assumption is less artificial as it seems.
We recall that each agent interacts with other agents randomly chosen from the whole population.
This already provides information  about the global level of cooperation, $f$.
Hence, at the global level, information about the distribution of strategies and of payoffs can assumed to be public knowledge.
And estimating (or observing) which of two subpopulations is wealthier might be easier to achieve than calculating an individual payoff. 

\begin{figure}[htbp]
  \centering
  \includegraphics[width=0.45\textwidth]{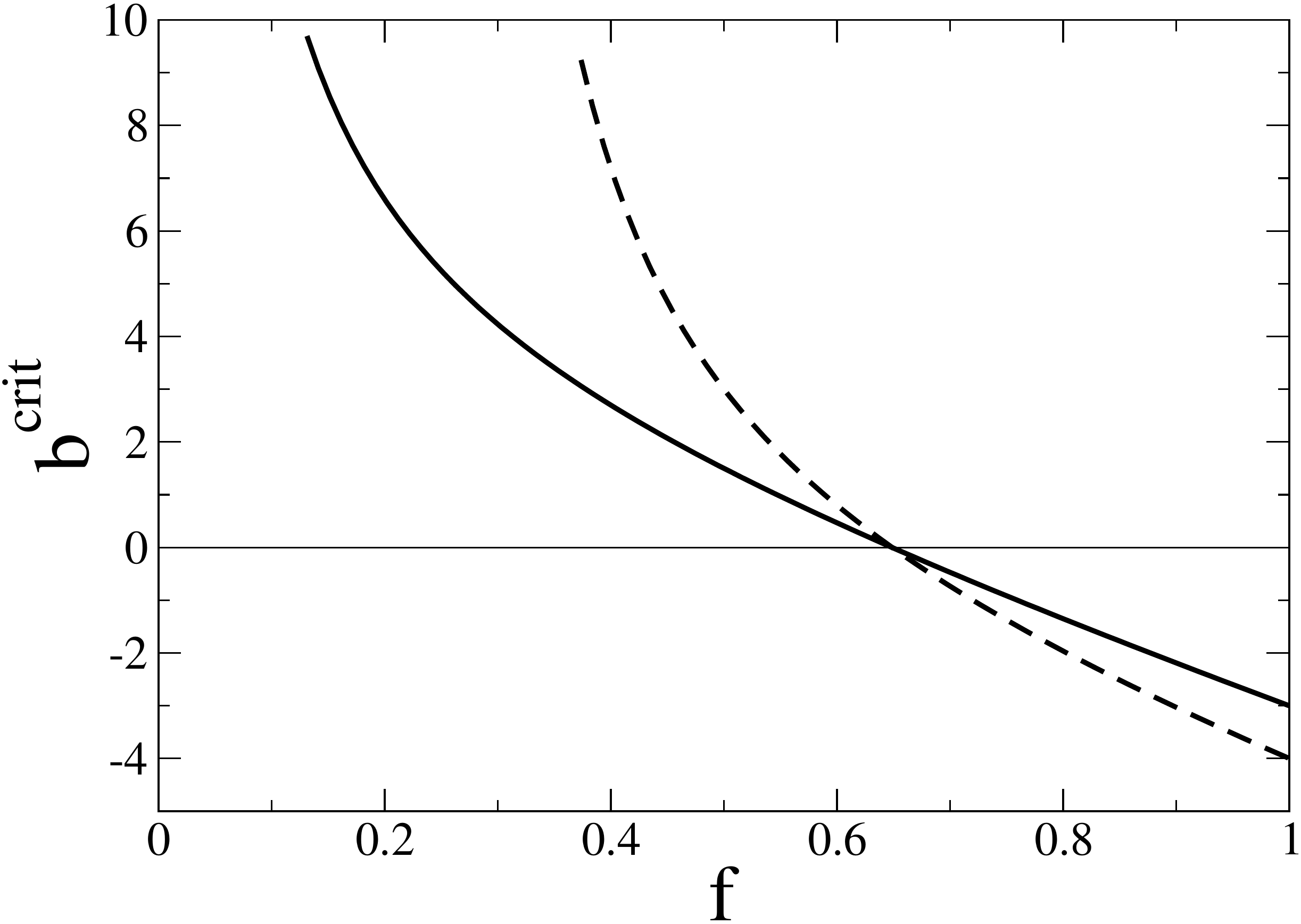}
  \caption{Critical bonus level, Eqn.~\eqref{eq:21-new} dependent on the fraction of cooperators, $f$. Solid line: $p$=1, dashed line: $p$=0.75.}
  \label{fig:bonus2}
\end{figure}

With these considerations in mind we modify the decision rule of Eqn.~\eqref{eq:13} as follows:
\begin{align}
  \label{eq:13-new}
q(s_{i}|(1-s_{i}),f)
  = \frac{\exp{\{\beta g_{i}(s_{i},f)\, f\}}}{
\exp{\{\beta g_{i}(s_{i},f)\, f\}}+\exp{\{\beta g_{i}(1-s_{i},f)[1-f]\}}}
\end{align}
The modified transition rates impact the odds ratio, Eqn.~\eqref{eq:14}, which eventually changes the condition to determine the critical bonus level:
\begin{align}
  \label{eq:17-new}
 f \left\{R f+ S \left [1-f\right] + \frac{p}{n_{g}} b\right\} > [1-f] \left\{T f + P  \left [1-f\right] + \frac{(1-p)}{n_{g}} b\right\} 
\end{align}
Instead of Eqn.~\eqref{eq:21},  with  the canonical payoff values $\{T,R,P,S\}=\{5,3,1,0\}$ we now have:
\begin{align}
  \label{eq:21-new}
  \hat{b}^{\mathrm{crit}}(f,p)= \frac{n_{g}}{pf -(1-p)[1-f]}\left\{-7f^{2}+3f+1\right\}
\end{align}
This is a very different expression for the critical bonus, and a comparison between Figure~\ref{fig:bonus1} and Figure~\ref{fig:bonus2} shows a very different dependence on $f$. 

Let us first consider again $p=1$, which results in
\begin{align}
  \label{eq:222-new}
  \hat{b}^{\mathrm{crit}}(f,p=1)= n_{g} \left[-7f+3+1/f\right]
\end{align}
This function diverges for $f\to 0$, i.e. without an small initial fraction of cooperators, no transition to cooperation can be induced.
With increasing $f$, the critical bonus level instead decreases, and even reaches zero at $\hat{f}^{\mathrm{crit}}=0.643$.

Considering Eqn.~\eqref{eq:21-new} for the case $p<1$ would not improve the situation, as Figure~\ref{fig:bonus2} shows. 
For all values $p<1$, the denominator decreases, this way raising the level of the critical bonus.
We still find the same value $\hat{f}^{\mathrm{crit}}$ for the vanishing critical bonus.
But for $f<\hat{f}^{\mathrm{crit}}$, the critical bonus is much higher because of the mistakes in identifying cooperators correctly.
Even if the functional dependence is similar, to start a trend towards increasing cooperation would need a much larger initial fraction of cooperators. 

This result has two implications.
First, if the cold start problem can be solved by choosing $f(0)>0$, e.g. 10\% of initial cooperators for large $p\to 1$, every bonus reward for cooperation \emph{improves} the situation, by increasing $f$ and this way decreasing the critical bonus needed at the next time step.
This is a virtuous cycle.  
Secondly, once a critical level $\hat{f}^{\mathrm{crit}}$ is reached, the government does not need to pay any bonuses for cooperation.
Instead, it can save this amount to increase its own wealth.
Hence, we reach a situation in which cooperation is sustained and the wealth, both of individual agents and the central authority, continuously increases.

\subsection{Governmental budget}
\label{sec:administrative-costs-1}

The positive conclusions drawn from the modified model can now be verified for the budget of the government which results from this redistribution mechanism.
We recall that in the model with local information, there was a (small) optimal level of cooperation $f^{\mathrm{opt}}$ to maximize the budget of the government, because higher levels of cooperation are associated with higher bonus payments.

The upper limit for the budget is still given by Eqn.~\eqref{eq:37}.
But instead of the critical bonus level $b^{\mathrm{crit}}$, Eqn.~\eqref{eq:21}, we now have to consider the new expression $\hat{b}^{\mathrm{crit}}$, Eqn.~\eqref{eq:21-new}, which results in:
\begin{align}
  \label{eq:38-new}
  \Delta W \leq \alpha N n_{g}\left\{-f^{2}+3f+1\right\}
  - (1-\alpha) N n_{g} \{-7f^{2}+3f+1\} \frac{\left\{pf + (1-p) [1-f]\right\}}{\left\{pf - (1-p) [1-f]\right\}}
\end{align}
\begin{figure}[htbp]
  \centering
  \includegraphics[width=0.45\textwidth]{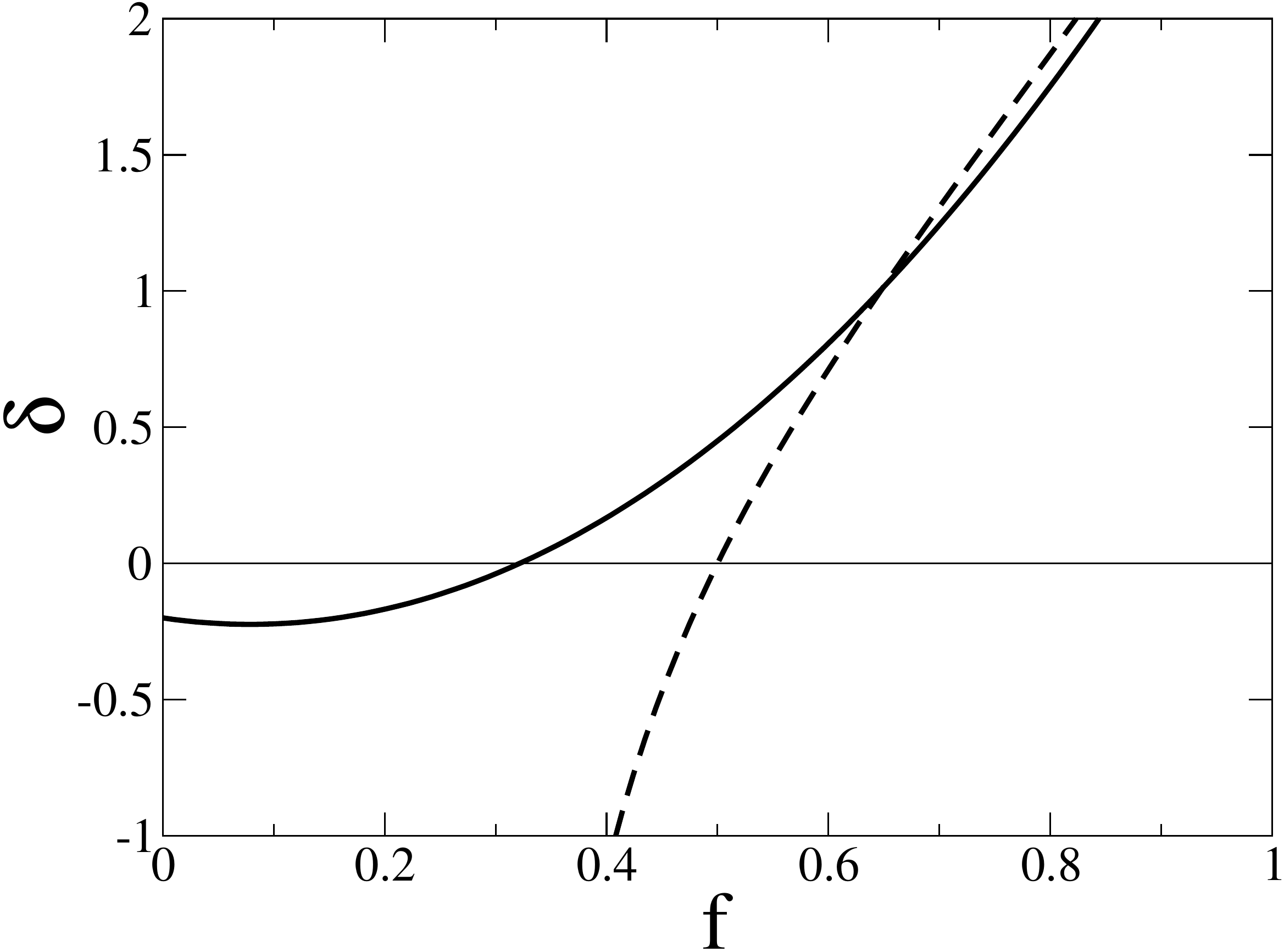}(a)
  \hfill
   \includegraphics[width=0.45\textwidth]{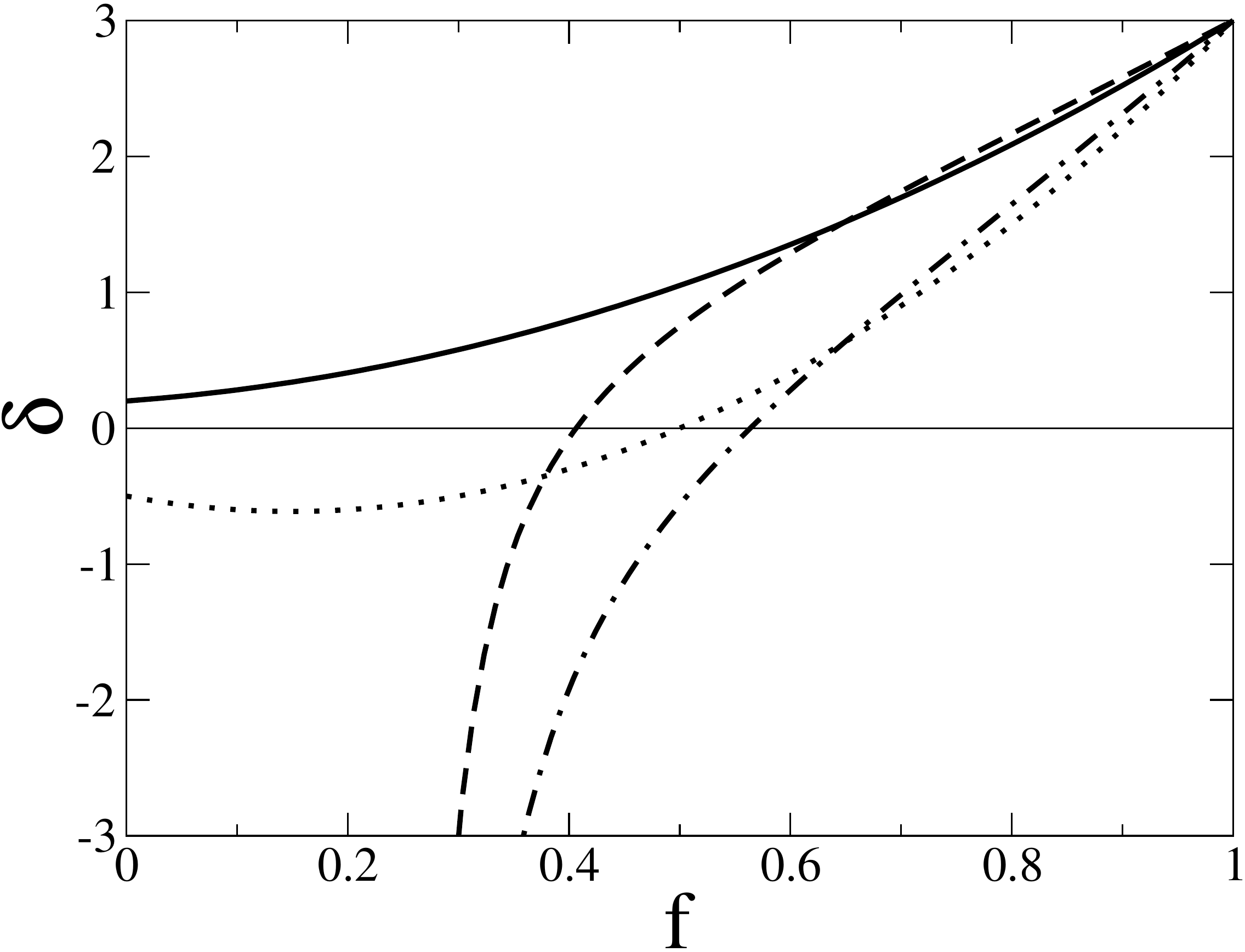}(b)
  \caption{Budget level, Eqn.~\eqref{eq:38-new} dependent on the fraction of cooperators, $f$. Solid line: $p$=1, dashed line: $p$=0.75. Different tax levels: (a) $\alpha$=0.4, (b) $\alpha$=0.25 (dotted/dashed-dotted), $\alpha$=0.60 (solid/dashed). } 
  \label{fig:costs2}
\end{figure}

This equation is not easily simplified, therefore in Figure~\ref{fig:costs2} we plot it for different levels of $\alpha$ and $p$.
Let us first discuss again the case $p=1$ which gives:
\begin{align}
  \label{eq:388}
  \delta(f,\alpha,p=1)= \frac{\Delta W}{N n_{g}} \leq -f^{2}(8\alpha-7) + f(6\alpha -3) +(2\alpha-1) 
\end{align}
As Figure~\ref{fig:costs2}(a) shows, a tax level $\alpha$=0.4 is not sufficient for the government to pay a bonus if the fraction of cooperators is low.
However, if $f$ could be increased above a critical level $\bar{f}^{\mathrm{crit}}$, which follows from $\delta(f,\alpha,p)$=0, the governmental budget becomes positive and quickly increases with $f$.
This will cover the administrative costs, $C$, and may even allow for savings, $S$.

Comparing Figures~\ref{fig:bonus2}, \ref{fig:costs2} we note that $\bar{f}^{\mathrm{crit}}<\hat{f}^{\mathrm{crit}}$.
That means, for a level of cooperation $\bar{f}^{\mathrm{crit}}\leq f \leq \hat{f}^{\mathrm{crit}}$ the government has to pay bonuses to reward cooperating agents.
However, these bonuses are small enough to allow a positive budget $\delta$.

Further comparing the cases $p=1$ and $p<1$, we observe that for $p<1$ the critical level 
$\bar{f}^{\mathrm{crit}}$ for positive budgets is considerably higher. 
At high values of $f$, there is an intersection of the two curves, which indicates that a budget with $p<1$ would become higher than a budget with $p=1$.
This intersection, however, only denotes the critical fraction $\hat{f}^{\mathrm{crit}}$ at which the critical bonus level becomes negative (compare Figures~\ref{fig:bonus2} and \ref{fig:costs2}). So, for $f\geq \hat{f}^{\mathrm{crit}}$ the government does not need to reward cooperation anymore, and therefore can abandon the effort to identify cooperators.
This does \emph{not} imply that all agents are already cooperating, in fact only 2/3 do so.
The remaining defectors will be convinced to switch to cooperation simply by comparing the total wealth of the two subpopulations.
Thus, nudging or rewards for cooperation are no longer needed.

\begin{figure}[htbp]
\includegraphics[width=0.45\textwidth]{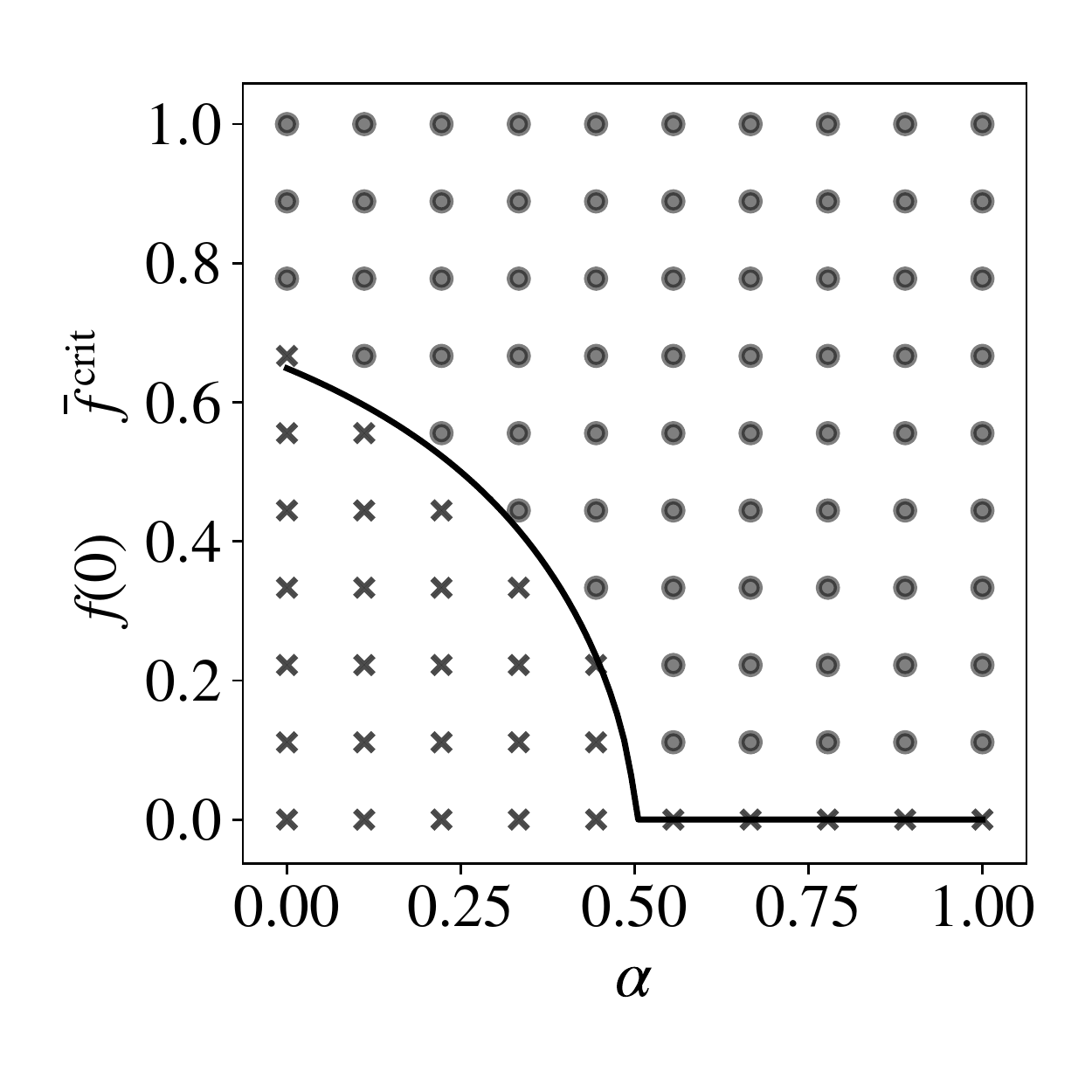}(a)
  \hfill
  \includegraphics[width=0.45\textwidth]{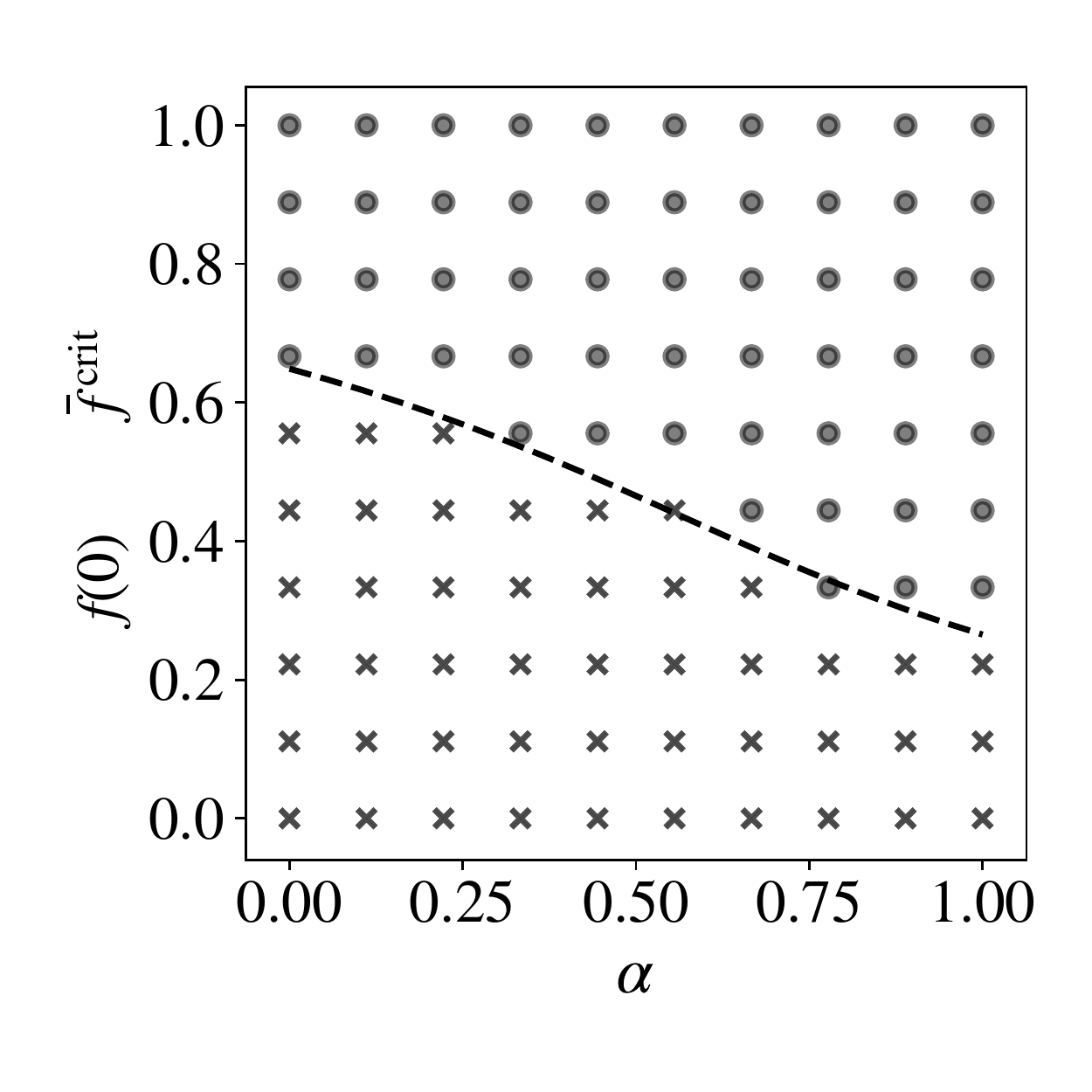}(b)
  \caption{
    Phase plot $(x,y)=(\alpha,f(0))$ for the final fraction of collaborators $f^{\mathrm{stat}}\approx 1$ (full dots) and   $f^{\mathrm{stat}}\approx0$   (crosses). (a) $p$=1, (b) $p$=0.75. (dashed lines) expected $\bar{f}^{\mathrm{crit}}$, Eqn.~\eqref{eq:388}.
    }
  \label{fig:sim2}
\end{figure}

To validate the above results and to further illustrate the cold-start problem, we have performed agent-based simulations also for this second scheme of global information.
The results are shown in Figure~\ref{fig:sim2} as a phase plot with  different tax rates $\alpha$ on the $x$-axis and different initial fractions of cooperators $f(0)$ on the $y$-axis.
The symbols in this phase plot encode the stationary fraction of cooperators, $f^{\mathrm{stat}}$, as the outcome of our simulations. 
We observe only two possible values, either full cooperation, $f^{\mathrm{stat}}\approx 1$ or no cooperation $f^{\mathrm{stat}}\approx 0$.
The transition between these two ``phases'' is denoted by critical threshold,  $\bar{f}^{\mathrm{crit}}$ from  Eqn.~\eqref{eq:388}.
For $f>\hat{f}^{\mathrm{crit}}$, agents start to perceive  cooperation as the best strategy.

Because for $f>\bar{f}^{\mathrm{crit}}$ we  always end up in full collaborations,
the government can lower the tax rate $\alpha$ once this critical level is reached.
Moreover, it can also stop to pay bonuses.
For $p$=0.75, the results are similar to $p$=1, but the phase of no cooperation is clearly extended.
In particular, full collaborations can no longer be achieved, regardless of the tax level, if the initial fraction of collaborators is too low.
This is in line with the findings shown in  Figure~\ref{fig:bonus2}.

In Figure~\ref{fig:sim2} we note the good agreement between the simulations and the analytical prediction for the  two different values of $p$.
But the match cannot be perfect because for $\beta<\infty$, we have stochastic fluctuations  and, hence, some defectors will not switch to cooperation and cooperators will switch to defection.
Thus, the simulated threshold value for $\hat{f}^{\mathrm{crit}}$ has to be larger than the analytical one, given by Eqn.~\eqref{eq:388}, and full co-operation ($f=1$) is never exactly reached.

\section{Discussion}
\label{sec:discussion-1}

The guiding question of our investigations, namely ``Should the government reward cooperation?'' can now be answered in a deliberative manner: ``Yes, if it can afford it.''
We have identified critical constellations in which it simply can't.
One limiting factor is the tax rate $\alpha$, which has to be sufficiently high to collect enough taxes to pay bonuses for cooperators.
A second limiting factor is the accuracy $p$ to identify those agents that would deserve a bonus for their cooperation.
Already small inaccuracies lead to a considerable deterioration, so it pays off to invest in high levels of $p$. 

As our agent-based model has demonstrated, governmental effort can become much more efficient if it is combined with the right information scheme to inform an agent's decisions.
The local information scheme, in which only individual gains are compared, definitely has its limitations.
The government is able to induce a certain level of cooperation, by simply paying the payoff differences in form of a bonus.
However, this cooperation is not sustainable, because it clearly depends on the governmental payments.
These become very costly for high levels of cooperation, to compensate for the strong incentives to defect in a cooperating environment.

The global information scheme, which was presented as an alternative for the decision rule, has the advantage that it leads to a sustainable cooperation.
Precisely, once a critical level of initial cooperation is reached, for which we present an analytical solution, no additional bonuses from the government is required to increase cooperation.
This requires to solve the ``cold-start problem'' to seed initial cooperation.
Governmental bonuses need to be very high if the level of cooperation is low.
As our model suggests, one way to overcome this daunting period would be a high initial tax rate, which would allow to pay these bonuses.
We found for the model that $\alpha\geq 0.5$ would be sufficient.
Once a certain level of cooperation is reached, $\alpha$ could be lowered, i.e. the tax rate could become a decreasing function of the fraction of cooperators, $\alpha(f)$.

Still taxes are needed to support the administrative tasks of the government, i.e. tax collection, identifying cooperators, and paying bonuses.
In our paper, we have not explicitly discussed these costs, $C$, although we have assumed that they shall be covered by the governmental budget, $\Delta W$, which has to be much higher than the limit case, $C=0$, used for the calculations.
Our investigations hold also if arbitrary floor values $C>0$ are chosen.

In the global information regime, 
we have noticed that the task of identifying cooperators is particularly important for  low levels of cooperation.
At the same time, this effort is also very costly.
Hence, we could assume that the accuracy level becomes a decreasing function of the fraction of cooperators, $p(f)$.
It starts with $p\to 1$, this way consuming most of the administrative budget, i.e. $p(C)$.
For larger $f$, the value of $p$ and hence the fraction of the costs spent on identifying cooperators can be lowered. 
This discussion is continued in a subsequent paper.

In conclusion, our agent-based model reflects fundamental mechanisms of wealth redistribution via the tax collection of a central authority.
Wealth originates from the interactions of agents, for which we have assumed a game-theoretical framework.
Differently from classical games, like the prisoner's dilemma, defectors may not get the highest gain, because the government uses part of the collected taxes to reward cooperators with a special bonus.
This bonus has to be large enough to make cooperation at least as attractive as defection.
Our game-theoretical setting allows us to calculate this critical level, for two different decision rules.

Our abstract discussion omits the question whether the government shall intervene this way.
Rewarding agents that contribute to a common good or cooperate under defecting conditions surely generates a positive impact on the level of cooperation. 
With our paper we have contributed to the interesting problem whether that is also a \emph{feasible scenario}.
We could quantify the conditions for that.
The most important insight is, perhaps, that the monetary incentives don't play the major role to achieve collaboration in a sustainable manner.
Instead, one needs to provide the right information to agents to allow them to make the right decision.

\setlength{\bibsep}{1pt}

\end{document}